\documentclass{optica-article}

\journal{optcon}

\articletype{Research Article}
\newcommand\numberthis{\addtocounter{equation}{1}\tag{\theequation}}

\begin{document}

\title{Faraday rotation signal amplification using high-power lasers}

\author{P.-A. Gourdain,\authormark{1,2,*}, A. Bachmann\authormark{1},
I. N. Erez\authormark{1,2}, M. E. Evans\authormark{1,2}, F. Garrett\authormark{1}, J. Hraki\authormark{1}, H. R. Hasson\authormark{1,2}, S. McGaffigan\authormark{1}, I. West-Abdallah\authormark{1,2}, J. R. Young\authormark{1,2}
}

\address{\authormark{1}Department of Physics and Astronomy, University of Rochester, Rochester NY 14627, USA\newline
\authormark{2}Laboratory for Laser Energetics, University of Rochester, Rochester NY 14627, USA}

\email{\authormark{*}gourdain@pas.rochester.edu} 



\begin{abstract}
Magnetic fields play an important role in plasma dynamics, yet it is a quantity difficult to measure accurately with physical probes, whose presence disturbs the very field they measure.  The Faraday rotation of a polarized beam of light provides a mechanism to measure the magnetic field without disturbing the dynamics, and has been used with great success in astrophysics and high energy density plasma science, where physical probes cannot be used. However, the rotation is typically small, which degrades the accuracy of the measurement. Paradoxically, the main source of error is the probe beam itself. Since polarization cannot be measured directly, detectors rely on a polarizer to measure a small change in beam intensity instead.  In this work, we show how suppress the beam intensity that is not part of the Faraday rotation signal by taking forming an optical derivative. Since the rotation measurement is now strictly proportional to the beam intensity, the system allows to amplify the rotation measurement simply by increasing the laser power.
\end{abstract}
\section{Introduction}

Faraday rotation (FR)\cite{stone1964radiation}, the change of light polarization caused by a magneto-ionized gas, thereafter called a plasma, has found applications in a wide range of disciplines over the years, from laser engineering \cite{aplet1964faraday} to astrophysics\cite{zeldovich1983magnetic}. The landmark experiment, done by Michael Faraday in 1845\cite{crowther1918life}, showed that the polarization of an electromagnetic wave can be affected by a magnetic field parallel to the direction of propagation. The magnetic field triggers a birefringence inside the material, which in turn changes the polarization angle $\theta_{FR}$ as
\begin{equation*}
    \theta_{FR}=VBL,
\end{equation*}
where $B$ is the magnetic field strength, projected along the direction of propagation of the electromagnetic wave, $L$ is the propagation length in the birefringent medium, and $V$ is the Verdet constant. A similar effect caused by magnetic fields transverse to the direction of propagation is called the Cotton-Mouton effect. While the Verdet constant can be difficult to compute for materials, it is relatively straightforward for a plasma\cite{pisarczyk1990sarkisov}, 
\begin{equation}\label{eq:FR_formula}
    \theta_{FR}(x,y)=\frac{e^3\lambda^2}{8\pi^2\epsilon_0m_e^2c^3}\int_L{n_e(x,y,z)\mathbf B(x,y,z)\cdot d\mathbf{z}},
\end{equation}
where $e$ is the fundamental charge, $\lambda$ is the laser wavelength, $\epsilon_0$ the vacuum permittivity, $m_e$ the electron mass, $n_e$ the plasma electron number density. We supposed here that the light travels along the $z$-axis of our coordinate system, the $x$-axis being horizontal and the $y$-axis vertical. This dependence allows to compute the magnetic field $B_z$, provided that the electron density $n_e$ is already known. This effect has been used intensively to measure the magnetic field strength of the interstellar medium\cite{gray1998large} by using background polarized sources, generating synchrotron radiations across a variety of wavelengths in the radio range\cite{manchester1972pulsar}. FR was also used actively to highlight how bipolar jets spawned from accretion disks\cite{bachiller1996bipolar} are sensitive to the Hall effect\cite{konigl2010}, an effect \cite{gourdain2013impact} also found in high energy density plasma jets produced in the laboratory\cite{suzuki2009formation,gourdain2010initial}. 

However, using FR to measure magnetic fields in the laboratory has proven challenging when using a visible laser\cite{swadling2014diagnosing} rather than microwaves\cite{zhang2010interaction}. The laser wavelength is a scaling factor in front of the integral of Eq. \eqref{eq:FR_formula}. As the wavelength becomes smaller, the cut-off density becomes larger and deeper regions of the plasma can be probed\cite{swadling2014diagnosing}. However, the Faraday rotation angle is now greatly reduced and becomes difficult to measure. In practice, this scaling factor acts as a attenuation factor.

In this paper we show that an S-waveplate combined with a neutral density filter with radial variation scan be used to amplify the Faraday rotation signal. In general, the S-waveplate\cite{beresna2011radially} utilizes polarization to convert a linearly polarized Gaussian beam into a radially or azimuthally polarized beam\cite{bauer2014nanointerferometric}. It is the complementary of the vortex plate\cite{Zhan:09,Galvez:12,boyd2016quantum}, which  affects the phase of the light rather than its polarization. The vortex plate achieves this by using a helical cavity with increasing depth corresponding to the azimuthal angle of the plate, changing the incoming phase of light azimuthally from 0 to $2m\pi$, where $m$ is the plate charge. In contrast, the S-waveplate uses nanogratings with their alignment rotating with the plate's azimuthal angle to change the incoming polarization azimuthally from 0 to $2m\pi$, also using $m$ as the plate charge. The proposed system gives the spatial derivative of the FR rotation signal, \textit{de facto} removing the bias light from the probe beam. With the bias gone, the laser intensity can be increased, improving the signal-to-noise ratio of the measurement.

\section{The optical derivative of a polarized beam}
\begin{figure}[ht]
\centering
    \includegraphics[width=4.5in,trim={0 0 0 0}, clip]{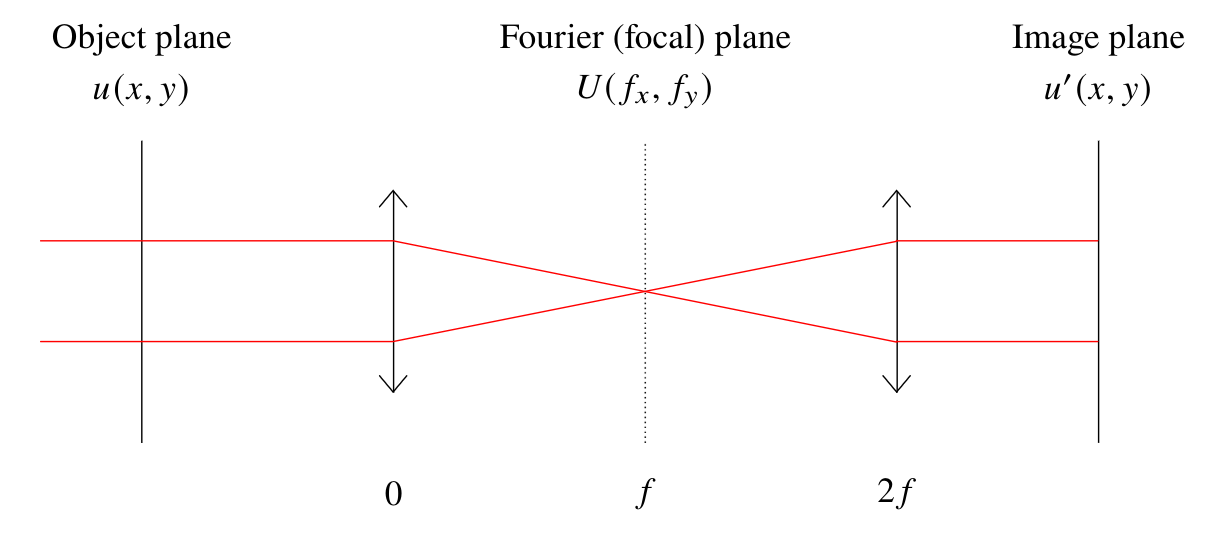}
\caption{Two-$f$ system uses two identical lenses of focal length $f$. The $z$-axis is along the optical axis.}
\label{fig:2f_setup}
\end{figure}

When a light beam travels through the experimental setup shown in Fig. \ref{fig:2f_setup}, its spatial constituents undergo a transformation into its spectral (i.e. spatial frequency) constituents at the focal plane of the initial lens. This procedure finds mathematical expression through the utilization of the Fourier transform:
\begin{equation*}\label{eq:fourier_theorem}
    U(f_x,f_y)=\int_{-\infty}^{+\infty}\int_{-\infty}^{+\infty}u(x,y)e^{-i2\pi x f_x}e^{-i2\pi y f_y}dxdy,
\end{equation*}
where $U(f_x,f_y)$ is the intensity representing the Fourier transform of the light on the focal plane of the lens\cite{Goodman2017-eo}.

\subsection{Description of the S-waveplate using Jones matrix}
The description of the S-waveplate can be effectively realized through the application of Jones matrices\cite{theocaris1979matrix,roux2006geometric}. This element can be divided into a retarder introducing a phase delay of $\varphi=\pi$ and a rotation matrix that imparts a rotation determined by the orientation $\phi$ of the nanogratings. Mathematically, the explicit expression for $\mathscr S$ takes the form:
\begin{equation*}
\mathscr S= \text{e}^{\frac{-i\varphi}{2}}\left[{\begin{array}{*{20}{c}}{{{\cos}^2}\phi + {e^{i\varphi}}{{\sin}^2}\phi}&{({1 - {e^{i\varphi}}} )\cos \phi \sin \phi}\\{({1 - {e^{i\varphi}}} )\cos \phi \sin \phi}&{{e^{i\varphi}}{{\cos}^2}\phi + {{\sin}^2}\phi}\end{array}} \right].
\end{equation*}
Given that $kR=\pi$, the expression can be simplified to:
\begin{equation}\label{eq:jones_matrix}
\mathscr S(x,y) =-i \left[{\begin{array}{*{20}{c}}{\cos \Theta (x,y)}&{\sin \Theta (x,y)}\\{\sin \Theta (x,y)}&{-\! \cos \Theta (x,y)}\end{array}} \right],
\end{equation}
where the local polarizing angle $\Theta(x,y)$ is twice the nanograting angle $\phi(x,y)$.

\subsection{Beam with variable intensity and constant linear polarization}
We start with a linearly polarized laser beam having a stable polarization angle $\theta$ relative to the polarization of the S-plate. Any electric field with a stable polarization in the object plane can be expanded into a discrete summation of sine and cosine functions using the discrete Fourier transform as 
\begin{equation}
\mathbf E(x,y)=\frac{\mathbf A_0}{2}+\sum_{n=1}^{+\infty}\cos(\textbf k_n\cdot\textbf r)\mathbf A_n+\sin(\textbf k_n\cdot\textbf r)\mathbf B_n,
\label{eq:Fourier_decomposition}
\end{equation}
where $\textbf k_n\cdot \textbf r=k_{x,n}x+k_{y,n}y$, $\mathbf A_n=A_n\cos\theta\,\mathbf x+A_n\sin\theta\,\mathbf y$, and $\mathbf B_n=B_n\cos\theta\,\mathbf x+B_n\sin\theta\,\mathbf y$. Note that $A_n$ and $B_n$ can be complex. To illustrate the operation done in the Fourier plane by the S-waveplate, we examine each component of $\mathbf E(x,y)$ individually. We denote the component of $\mathbf E$ corresponding to $\mathbf k \in\{\mathbf k_1,\hdots,\mathbf k_n,\hdots,+\infty\}$ as $\mathbf E_{\textbf k}(x,y)=\mathbf A_{\textbf k}\cos(k_xx+k_yy)+\mathbf B_{\textbf k}\sin(k_xx+k_yy)$. The Fourier transform of $\mathbf E_\textbf k$ is given by
\newcolumntype{K}{>{\raggedleft\arraybackslash$}p{6cm}<{$}}
\begin{align*}
   \mathscr E_{\textbf k}(f_x,f_y)=&\left[\pi\delta\left(\frac{k_x}{2\pi}-f_x\right)\delta\left(\frac{k_y}{2\pi}-f_y\right)+\pi\delta\left(\frac{k_x}{2\pi}+f_x\right)\delta\left(\frac{k_y}{2\pi}+f_y\right)\right]\mathbf A_{\textbf k}\\
    +\frac{1}{i}&\left[\pi\delta\left(\frac{k_x}{2\pi}-f_x\right)\delta\left(\frac{k_y}{2\pi}-f_y\right)-\pi\delta\left(\frac{k_x}{2\pi}+f_x\right)\delta\left(\frac{k_y}{2\pi}+f_y\right)\right]\mathbf B_{\textbf k}.
\end{align*}
When introducing an S-waveplate at the Fourier plane $(f_x,f_y)$ where the orientation of the local polarizing angle $\Theta$ varies azimuthally from 0 to $2\pi$, the Jones matrix $\mathscr S$ of Eq. \eqref{eq:jones_matrix} can be expressed in term of the frequencies $f_x$ and $f_y$ as
\begin{equation*}
\mathscr S(f_x,f_y) = -i \left[{\begin{array}{*{20}{c}}{\frac{f_x}{\sqrt{f_x^2+f_y^2}}}&{\frac{f_y}{\sqrt{f_x^2+f_y^2}}}\\{\frac{f_y}{\sqrt{f_x^2+f_y^2}}}&{-\frac{f_x}{\sqrt{f_x^2+f_y^2}}}\end{array}} \right].   
\end{equation*}
Now, we combine the S-waveplate with a neutral density filter whose transmission varies linearly with the radius $\rho$ and is defined as\cite{gourdain2023true}
\begin{equation}
    \rho(f_x,f_y)=\sqrt{f_x^2+f_y^2}.
    \label{eq:ND_profile}
\end{equation}
In practice, the transmission cannot exceed 1. While a scaling factor could be used, it is omitted here. The S-waveplate and the neutral density filter collectively constitute a new element denoted as $\mathscr T$, with a Jones matrix given by
\begin{equation*}\label{eq:transmission_matrix}
\mathscr T(f_x,f_y) = -i\left[{\begin{array}{*{20}{c}}{f_x}&{f_y}\\{f_y}&{-f_x}\end{array}} \right].   
\end{equation*}
Upon performing the inverse Fourier transform of $\mathscr E_{\mathbf k}\mathscr T$, we obtain on the image plane\cite{gourdain2023true}
\begin{equation}\label{eq:general_optical_derivative}
\mathbf E_{\textbf k}'(x,y)=-\left[
\begin{array}{*{20}{c}}
    {\cos\theta}&{\sin\theta}\\
     {-\sin\theta}&{\cos\theta}
\end{array}\right] \nabla \left[A_{\textbf k}\cos(\mathbf k\cdot\mathbf r)+B_{\textbf k}\sin(\mathbf k\cdot\mathbf r)\right]
\end{equation}
where $\nabla$ is the gradient operator. Since the main results of the paper focus on intensities, we will drop the minus sign in Eq. \eqref{eq:general_optical_derivative} from later equations. As a result, the total electric field on the image plane is given by
\begin{equation}\label{eq:compact_general_optical_derivative}
    \mathbf E'(x,y)=\nabla_{-\theta}E(x,y),
\end{equation}
where the electric field $E(x,y)$ is the electric field along the polarization direction $\theta$, i.e.
\begin{equation*}
E(x,y)=\frac{A_0}{2}+\sum_{n=1}^{+\infty}A_n\cos(\textbf k_n\cdot\textbf r)+B_n\sin(\textbf k_n\cdot\textbf r),
\end{equation*}
and $\nabla_{\theta}$ is the rotated gradient operator given by
$$
\nabla_{\theta}=\left[
\begin{array}{*{20}{c}}
    {\cos\theta}&{-\sin\theta}\\
     {\sin\theta}&{\cos\theta}
\end{array}\right]
\nabla\text{ or }
\nabla_{\theta}=\left[
\begin{array}{*{20}{c}}
    {\cos\theta\,\partial_x-\sin\theta\,\partial_y}\\
     {\sin\theta\,\partial_x+\cos\theta\,\partial_y}
\end{array}\right]
$$
The resulting electric field of Eq. \eqref{eq:compact_general_optical_derivative} is the gradient of the initial electric field along the polarization direction $\theta$. However, its polarization is along the $-\theta$ direction now. Note that for a beam with a polarization angle $\theta=0$ (polarization along the $x$-axis), $\nabla_\theta$ becomes the usual gradient operator $\nabla$. In this case, the S-waveplate/neutral density filter give the gradient along two perpendicular polarization axes, allowing to measure $\partial_xE$ and $\partial_yE$ independently. This is a departure from the vortex plate setup \cite{gourdain2023true} where the derivative along $x$ and $y$ were combined together. 

\subsection{Beam with variable intensity and polarization}
We again start with a linearly polarized laser beam with constant initial polarization angle $\theta_i$ with respect to the S-waveplate polarization. However, the electric field $E(x,y)$ has now acquired a variable polarization $\theta_{FR}(x,y)$ from a magnetoactive medium. We are dropping the coordinates $x$ and $y$ hereafter for the terms on the RHS of most equations. After exiting the medium, the electric field is given by 
\begin{equation}\label{eq:initial_E_field}
\mathbf E(x,y) = E\left[
\begin{array}{*{20}{c}}
{\cos\theta}\\{\sin\theta}
\end{array}
\right],
\end{equation}
where $\theta=\theta_i+\theta_{FR}$. Since a spatial change in polarization simply corresponds to a change in intensity of the $x$- and $y$-polarization components of the beam, we can use Eq. \eqref{eq:compact_general_optical_derivative} for each polarization  independently to find the electric field at the exit of the augmented two-$f$ system
\begin{equation*}
    \mathbf E'(x,y)=\nabla_0(E\cos\theta)+\nabla_{-\pi/2}(E\sin\theta),
\end{equation*}
which we can write explicitly as
\begin{equation}\label{eq:general_polarization_derivative}
\mathbf E'(x,y)=\left[
\begin{array}{*{20}{c}}
    {\partial_x\left(E\cos\theta\right)+\partial_y\left(E\sin\theta\right)}\\
     {\partial_y\left(E\cos\theta\right)-\partial_x\left(E\sin\theta\right)}
\end{array}\right]
\end{equation}
If $\theta$ is constant, then Eq. \eqref{eq:general_polarization_derivative} is simply Eq. \eqref{eq:compact_general_optical_derivative}. 

\section{Implementation of a Faraday rotation measurement}

\begin{figure}[ht]
\centering
    \includegraphics[width=4.5in,trim={0 0 0 0}, clip]{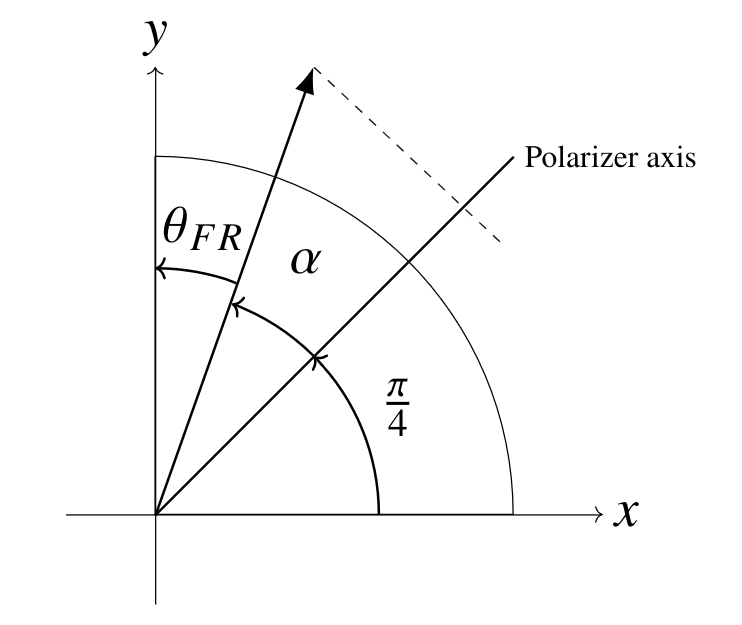}
\caption{Typical example of a Faraday Rotation measurement yielding an intensity that is proportional to $(1+\theta_{FR})$ using a polarizer that is rotated by $\pi/4$ compared to the initial laser beam polarization.}
\label{fig:FR_drawing}
\end{figure}
\subsection{Standard Faraday rotation measurement}\label{sec:standard_FR}
It is typical to start with a $y$-polarized (i.e. vertically polarized) beam. The beam then passes through the magnetoactive medium, impinging a Faraday rotation $\theta_{FR}$ to the polarization (see Fig. \ref{fig:FR_drawing}). At this point, a polarizer at $\pi/4$ from the vertical direction yields a new electric field 
\begin{equation}\label{eq:polarized_E_in}
    E_{\pi/4}(x,y)=E\cos(\alpha),
\end{equation}
where $\alpha=\pi/4-\theta_{FR}$ and $E(x,y)$ is the electric field strength. Now Eq. \eqref{eq:polarized_E_in} can be turned into
\begin{equation}\label{eq:intensity_cos_plus_sin}
    E_{\pi/4}(x,y)=\frac{1}{\sqrt 2}E\left[\cos(\theta_{FR})+\sin(\theta_{FR})\right].
\end{equation}
The intensity can be written as
\begin{equation}\label{eq:intensity_full}
    I_{\pi/4}(x,y)=\frac{1}{2}E^2\left[\cos^2(\theta_{FR})+2\cos(\theta_{FR})\sin(\theta_{FR})+\sin^2(\theta_{FR})\right].
\end{equation}
As $\theta_{FR}<<1$ in most low density magnetoactive media such as plasmas, we can do a Taylor expansion of Eq. \eqref{eq:intensity_full} up to the first order leading to 
\begin{equation}\label{eq:intensity_approx}
    I_{\pi/4}(x,y)\simeq\frac{1}{2}E^2\left(1+2\theta_{FR}\right).
\end{equation}
While this approach increases the measurement sensitivity compared to using a cross-polarizer downstream of the magnetoactive medium, the bias light (represented by the factor 1 in Eq. \eqref{eq:intensity_approx}) really reduces the signal-to-noise ratio. 

\subsection{Faraday rotation measurement with homogeneous beams}
We look first at the case where the laser intensity is homogeneous, i.e. $E(x,y)\simeq E_0$. We can choose $\theta_i=0$. In this case, the electric field of Eq. \eqref{eq:general_polarization_derivative} has a component along the $x$ given by
\begin{equation}\label{eq:FR_constant_E0_x}
    E'_x(x,y)=E_0\left(\partial_x\cos\theta_{FR}+\partial_y\sin\theta_{FR}\right)
\end{equation}
for the $x$-polarization and
\begin{equation}\label{eq:FR_constant_E0_y}
    E'_y(x,y)=E_0\left(\partial_y\cos\theta_{FR}-\partial_x\sin\theta_{FR}\right).
\end{equation}
for the $y$-polarization.
Since $\theta_{FR}<<1$ the zeroth order of Eq. \eqref{eq:FR_constant_E0_x} is
\begin{equation}\label{eq:E_field_FR_along_x}
    E'_x(x,y)\simeq E_0\partial_y\theta_{FR}
\end{equation}
In this case, the zeroth order of the intensity of the $x$-polarization is
\begin{equation}\label{eq:intensity_FR_along_x}
    I'_x(x,y)\simeq |E_0|^2{\partial_y\theta_{FR}}^2,
\end{equation}
Following the same line of thinking, Eq. \eqref{eq:FR_constant_E0_y} gives
\begin{equation}\label{eq:E_field_FR_along_y}
    E'_y(x,y)\simeq -E_0\partial_x\theta_{FR},
\end{equation}
with an intensity
\begin{equation}\label{eq:intensity_FR_along_y}
    I'_y(x,y)\simeq |E_0|^2{\partial_x\theta_{FR}}^2.
\end{equation}
We clearly see that the sensitivity of the measurement is reduced compared to Eq. \eqref{eq:intensity_approx} as we are looking at the square of the Faraday rotation angle derivative. However, while the Faraday rotation signal is still proportional to $E^2$ in Eqs. \eqref{eq:intensity_FR_along_x} and \eqref{eq:intensity_FR_along_y}, the bias signal of Eq. \eqref{eq:intensity_approx} has now disappeared. So the sensitivity can be recovered simply by increasing the power of the input laser.

\subsection{Faraday rotation measurement using beam shaping}
Most high power laser beams can have some degree of controlled heterogeneity \cite{dickey2018laser}. In this case, Eq. \eqref{eq:general_polarization_derivative} yields

\begin{equation}\label{eq:full_electric_field_after_2f_setup}
\mathbf E'(x,y)=E\left[
\begin{array}{*{20}{c}}
    {\cos\theta\partial_xE/E-\sin\theta\partial_x\theta+\sin\theta\partial_yE/E+\cos\theta\partial_y\theta}\\
     {\cos\theta\partial_yE/E-\sin\theta\partial_y\theta-\sin\theta\partial_xE/E-\cos\theta\partial_x\theta}
\end{array}\right].
\end{equation}

To recover the sensitivity obtained by the setup of Fig. \ref{fig:FR_drawing}, we take $\theta_i=0$. In this case, again supposing $\theta_{FR}<<1$, Eq. \eqref{eq:full_electric_field_after_2f_setup} simplifies to
\begin{equation}\label{eq:approx_full_electric_field_after_2f_setup}
    \mathbf{E}'(x,y)\simeq E\left[ \frac{\partial_xE}{E}+\partial_y\theta_{FR}\right]\mathbf{x}+E\left[\frac{\partial_yE}{E}-\partial_x\theta_{FR}\right]\mathbf{y},
\end{equation}
providing that $\partial_y\theta_{FR}/\theta_{FR}>>\partial_yE/E$. This condition is fulfilled when the magnitude of E is excessively large compared to its derivative. Using a polarizing beam splitter, we can record the intensity of the $x$-polarization given by
\begin{equation}\label{eq:high_sensitivity_x}
    I_x'(x,y)\simeq |E\partial_xE\left[ \partial_xE/E+2\partial_y\theta_{FR}\right]|,
\end{equation}
We see that the measurement of the Faraday rotation derivative is now similar to the standard measurement of section \ref{sec:standard_FR}. The bias is $\partial_xE/E$ rather than the "1" seen in Eq. \eqref{eq:intensity_approx} and the Faraday rotation has been replaced by its derivative. So, a high-power beam with small spatial dependence will amplify the Faraday rotation $\partial_y\theta_{FR}$ while reducing the inherent bias $\partial_xE/E$. Both will improve the signal-to-noise ratio substantially. Since the intensity of the $y$-polarization is
\begin{equation}\label{eq:high_sensitivity_y}
    I_y'(x,y)\simeq |E\partial_yE\left[ \partial_yE/E-2\partial_x\theta_{FR}\right]|,
\end{equation}
we can draw similar conclusions.
\section{Numerical simulations}
In the rest of the paper, we use ray tracing\cite{brea2019} with vector Rayleigh-Sommerfeld (VRS) diffraction \cite{shen2006fast} to compute the effect of each optical element on the intensity and polarization of Faraday rotation measurements. We will use two identical lenses with a diameter of 2 cm and a focal length of 25 cm.
\subsection{Quasi-homogeneous beam}
\begin{figure}[!htb]
    \centering
    \includegraphics[width=4.5in,trim={0 0 0 0}, clip]{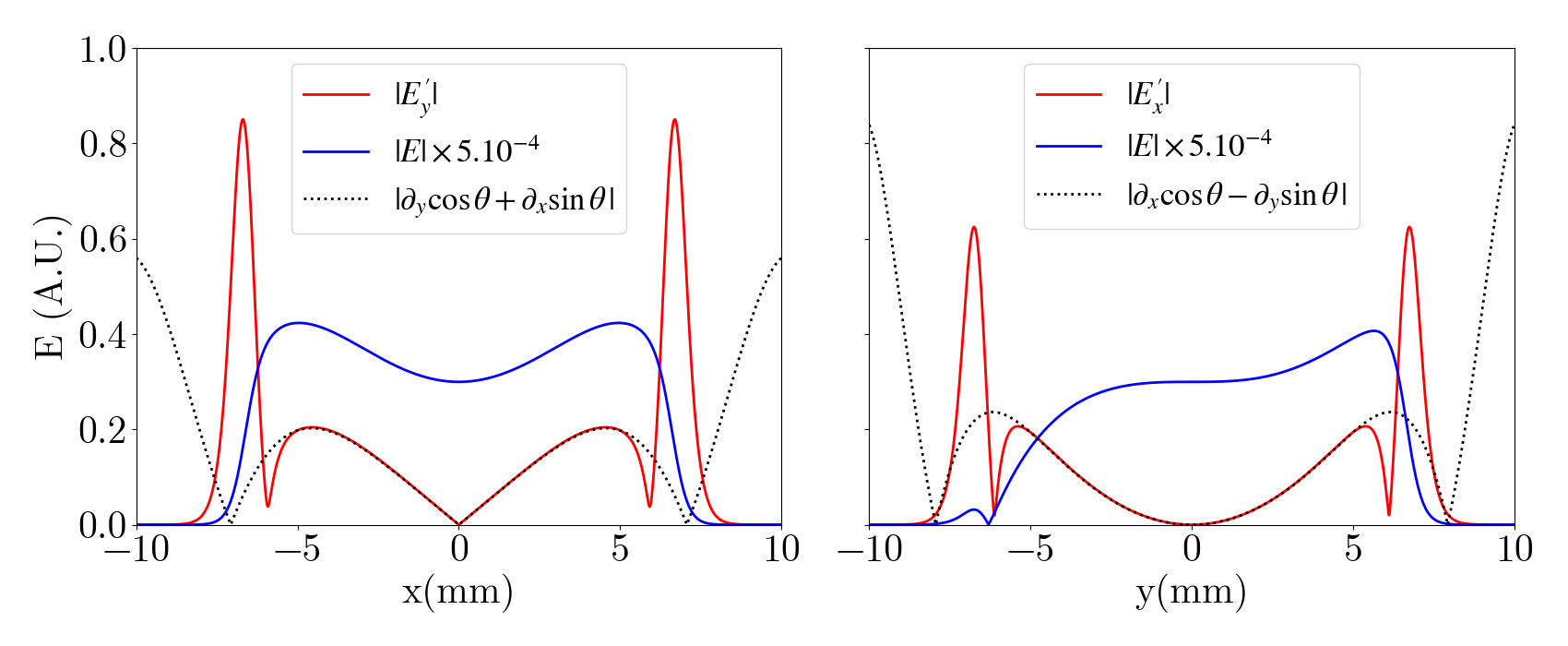}
     \caption{The electric field of the laser beam downstream of the setup of Fig \ref{fig:2f_setup} along the $x$-axis for the $y$-polarization (left) and along the $y$-axis for the $x$-polarization (right) for M=1 computed using VRS diffraction. The analytic values of the field are plotted using dotted lines. The magnitude of the initial electric field is indicated in blue. }
     \label{fig:constant_intensity_results_high_FR}
\end{figure}

\begin{figure}[!htb]
    \centering
    \includegraphics[width=4.5in,trim={0 0 0 0}, clip]{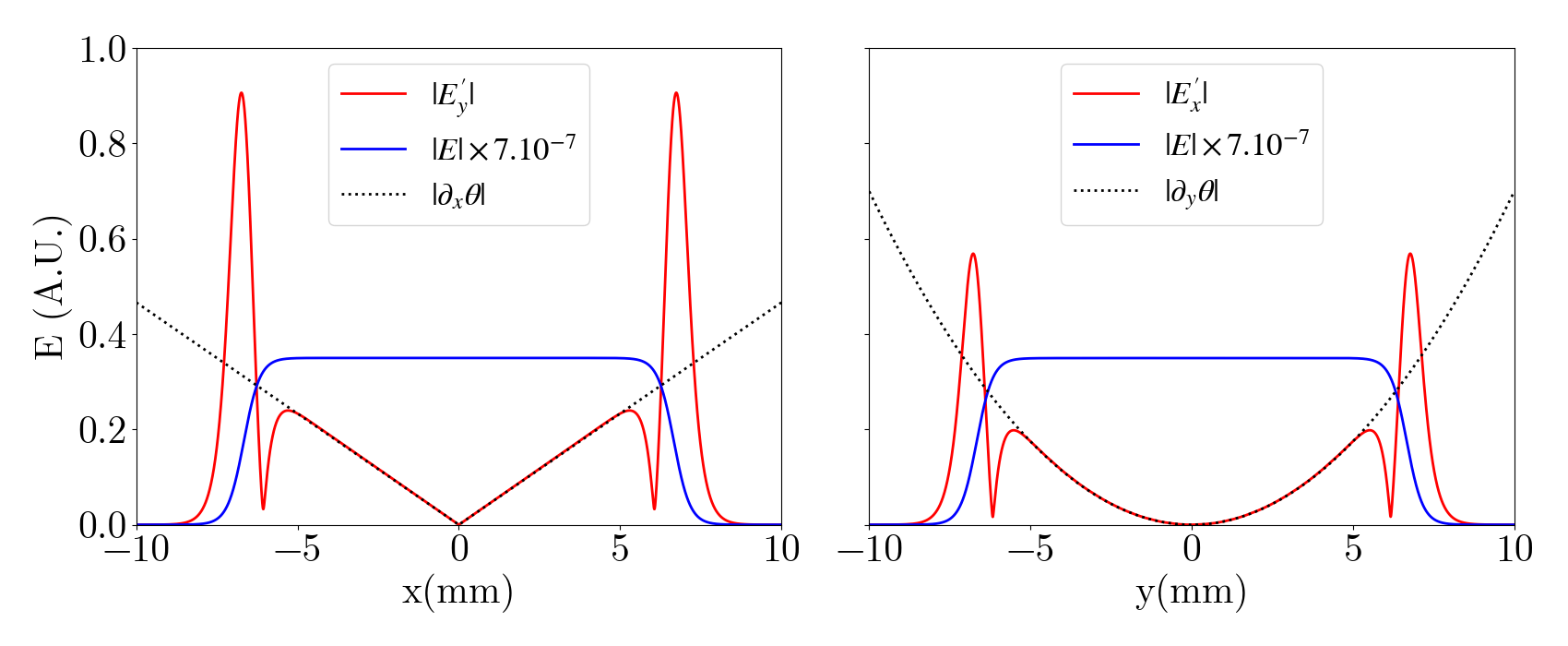}
     \caption{The electric field of the laser beam downstream of the setup of Fig \ref{fig:2f_setup} along the $x$-axis for the $y$-polarization (left) and along the $y$-axis for the $x$-polarization (right) computed using VRS diffraction. The analytic values are represented with dotted lines for small rotation angle(i.e. M=$10^{-3}$), they are approximately $\partial_x\theta_{FR}$ for the $y$-polarization (left) and $-\partial_y\theta_{FR}$ for the $x$-polarization (right). The magnitude of the initial electric field is indicated in blue.  }
     \label{fig:constant_intensity_results_low_FR}
\end{figure}

We use a laser beam profile in a large region of the ray-tracing domain that is mostly homogeneous and given by:
\begin{equation}\label{fig:E Field distribution}
    E(x,y)=\frac{E_0}{4}\left[1+\tanh\left(x_0-\frac{|x-x_c|}{\sigma}\right)\right]\left[1+\tanh\left(y_0-\frac{|y-y_c|}{\sigma}\right)\right],
\end{equation}
where $x_c$ and $y_c$ correspond to the beam center and $\sigma$ the transition width to 0. With the given parameters the full-width-half-maximum is 2$x_0$ along the $x$-axis and 2$y_0$ along the $y$-axis. Due to the effect of boundary conditions, we keep 2$x_0$ and 2$y_0$ on the order of 75\% of the computational domain width along the respective directions. The spatial dependence of E is shown in Figs. \ref{fig:constant_intensity_results_high_FR} and \ref{fig:constant_intensity_results_low_FR}.
The Faraday rotation given by
\begin{equation}\label{eq:FR_profile}
    \theta_{FR}(x,y,M)=M\pi\left(\left[\frac{x-x_{min}}{x_{max}-x_{min}}-\frac{1}{2}\right]^2+\left[\frac{y-y_{min}}{y_{max}-y_{min}}-\frac{1}{2}\right]^3\right),
\end{equation}
where the scaling factor $M$ allows to switch between large and small rotation, $x_{min}$ and $x_{max}$ are the domain limits along the $x$-axis, while $y_{min}$ and $y_{max}$ are the domain limits along the $y$-axis. This analytical expression will help to identify the FR feature unequivocally in the downstream beam and verify Eq. \eqref{eq:general_polarization_derivative}.

For M small (i.e. 1), the downstream electric field should follow Eqs. \eqref{eq:FR_constant_E0_x} and \eqref{eq:FR_constant_E0_y}, as plotted on Fig. \ref{fig:constant_intensity_results_high_FR}. For small rotation angle (i.e. M=$10^{-3}$) we expect a signal given by Eqs. \eqref{eq:E_field_FR_along_x} and \eqref{eq:E_field_FR_along_y}. Indeed we recover the linear dependence of the derivative along the $x$-axis, shown on the left hand side of Fig. \ref{fig:constant_intensity_results_low_FR}, and the quadratic dependence of the derivative along the $y$-axis, shown on the right hand side of Fig. \ref{fig:constant_intensity_results_low_FR}. 

Here we scaled the analytical value of the solution given by Eqs. \eqref{eq:FR_constant_E0_x} and \eqref{eq:FR_constant_E0_y} to the $x$-polarization of the output beam profile of Fig. \ref{fig:constant_intensity_results_high_FR} and  kept this scaling constant throughout the simulation section. 

\subsection{Shaped beam}
As Eqs. \eqref{eq:high_sensitivity_x}, \eqref{eq:high_sensitivity_y} show, it is possible to recover the sensitivity of the standard method given by Eq. \eqref{eq:intensity_approx}. However, this can only be achieved if the beam has well-defined spatial variations. In fact, these variations should be as simple as possible to allow for an easy measurement of the rotation. While we are using a laser for getting large intensities allowing to amplify the rotation signal, the coherence of the beam is not primordial as the S-waveplate affects the phase signal only by adding a constant. As a result, the beam profile can be controlled using engineered diffusers \cite{dickey2018laser} with an extremely high level of precision. We need to use here a diffuser that scrambles the phase rather than the polarization. 

\begin{figure}[!htb]
    \centering
    \includegraphics[width=4.5in,trim={0 0 0 0}, clip]{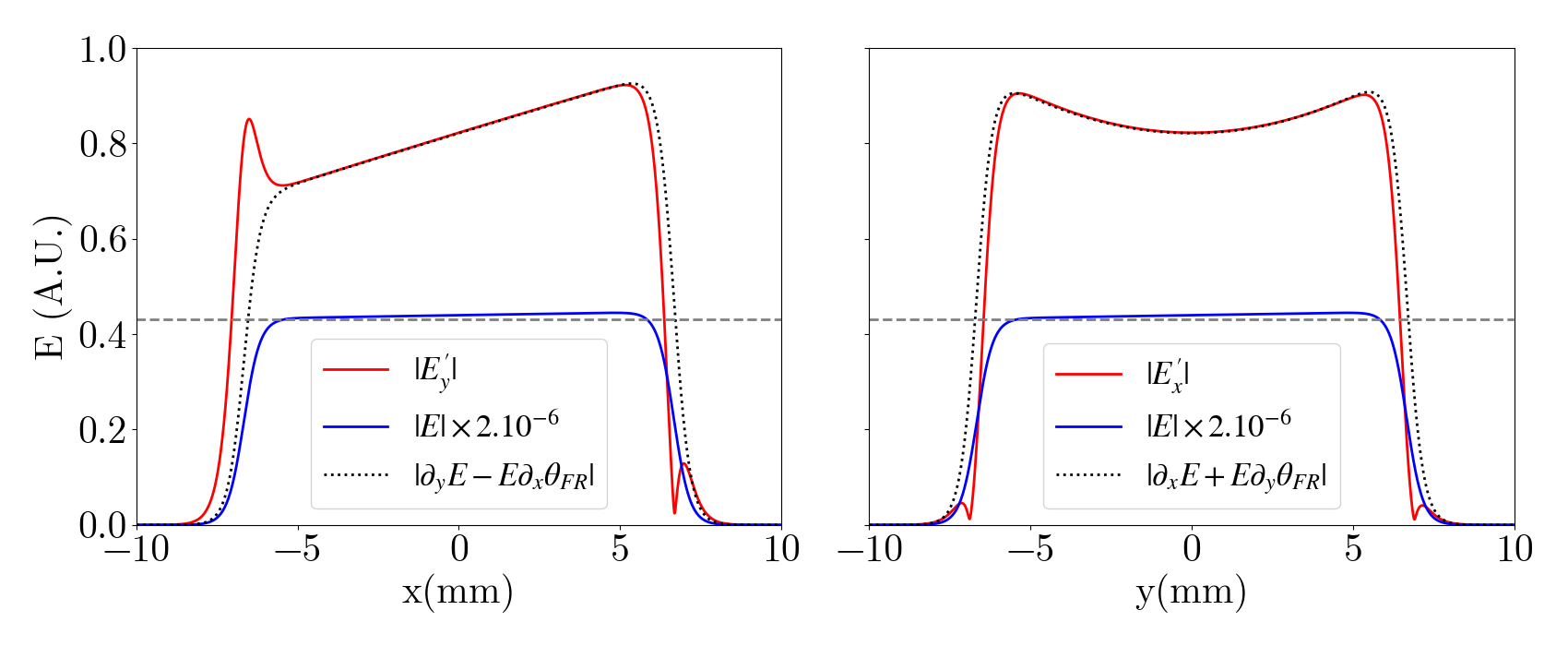}
     \caption{The electric field of the laser beam downstream of the setup of Fig \ref{fig:2f_setup} along the $x$-axis for the $y$-polarization (left) and along the $y$-axis for the $x$-polarization (right) computed using VRS diffraction. The analytic values are represented with dotted lines for a small rotation angle (i.e. M=$10^{-3}$). The initial electric field is indicated in blue. The horizontal dashed line are used to highlight the variation of the electric field of the initial beam.}
     \label{fig:variable_intensity_results_low_FR}
\end{figure}
To this end, we now add a small, linear variation to the previous beam, e.g.
\begin{align*}
    E(x,y)=&\frac{E_0}{4}\left[1+\tanh\left(x_0-\frac{|x-x_c|}{\sigma}\right)\right]\left[1+\tanh\left(y_0-\frac{|y-y_c|}{\sigma}\right)\right]\\
    &\times\left[|\alpha_xx+x_{bias}|+|\alpha_yy+y_{bias}|\right].\numberthis \label{fig:E Field distribution variation}
\end{align*}
If the variation is too large, then we could be back with a case similar to Eq. \eqref{eq:intensity_approx}, where the bias decreases the signal to noise ratio. However, if this is deemed necessary, we can reduce the impact on the Faraday rotation signal simply by increasing the base electric field. In this case the bias terms $\partial_xE/E$ and $\partial_yE/E$ in Eqs. \eqref{eq:high_sensitivity_x} and \eqref{eq:high_sensitivity_y} can be made small compared to the Faraday rotation signal, while the amplification terms $E\partial_xE$ and $E\partial_yE$ can be made large, \textit{de facto} amplifying the Faraday rotation signal. Using the same Faraday rotation profile of Eq. \eqref{eq:FR_profile}, Fig. \ref{fig:variable_intensity_results_low_FR} shows that ray tracing confirms Eq. \eqref{eq:approx_full_electric_field_after_2f_setup}. While not plotted, numerical simulations confirm both Eqs. \eqref{eq:intensity_FR_along_x} and \eqref{eq:intensity_FR_along_y}.    

\subsection{Noisy shaped beam}
It is now time to see what can be done about the beam noise. Clearly this is an open problem so far, at least for the low frequency noise. Yet, several techniques have shown good mitigation. For instance, low frequency noise can be removed using spectral dispersion \cite{skupsky1989improved,regan2000experimental} or micro-lens arrays \cite{zimmermann2007microlens,meng2021partially}. Polarization smoothing \cite{boehly1999reduction} cannot be used here since we are measuring polarization. 

\begin{figure}[!htb]
    \centering
    \includegraphics[width=4.5in,trim={0 0 0 0}, clip]{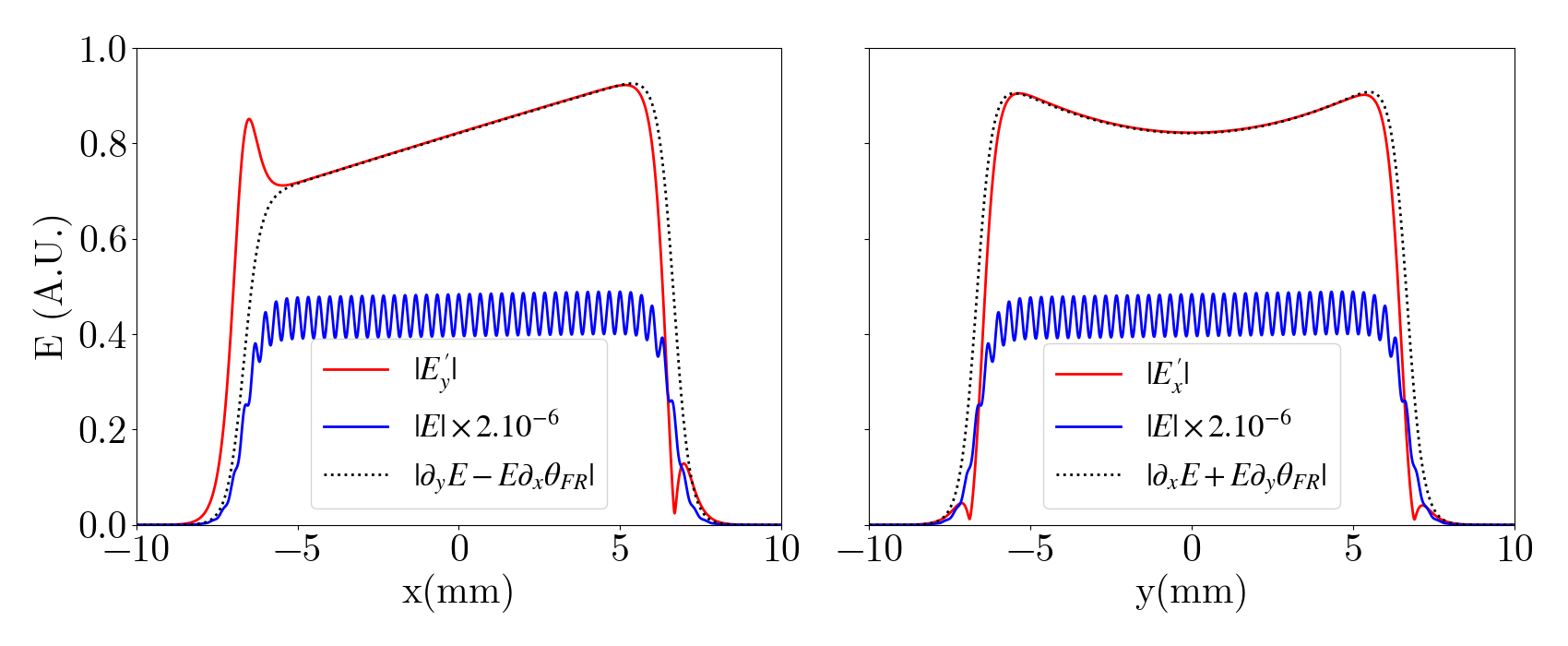}
     \caption{The electric field of the laser beam downstream of the setup of Fig \ref{fig:2f_setup} along the $x$-axis for the $y$-polarization (left) and along the $y$-axis for the $x$-polarization (right) using VRS. The expected values are plotted using dotted lines for a small rotation angle (i.e. M=$10^{-3}$). The initial electric field is indicated in blue and has a noise amplitude of 10\%. }
     \label{fig:variable_intensity_results_low_FR_with_noise}
\end{figure}

High frequency intensity noise can be removed more easily. This is simply done by placing an aperture in the Fourier plane of the setup shown in Fig. \ref{fig:2f_setup}\cite{Elias:52}. This can be combined directly inside the neutral density filter of Eq. \eqref{eq:ND_profile}, setting the transmission to zero at the filter periphery. Fig. \ref{fig:variable_intensity_results_low_FR_with_noise} shows that a high frequency noise up to 10\% has almost no impact on the Faraday rotation measurement, once filtered optically. 

\section{Conclusions}
This work shows that the combination of an S-waveplate with a linearly varying neutral density filter allows to compute the spatial derivative of the Faraday rotation angle. The $y$-polarization exiting the setup will carry the partial derivative along the $x$-axis, while the $x$-polarization will carry the partial derivative along $y$. In a standard setup, where the Faraday rotation signal appears on top of the beam carrier, increasing the beam energy does not necessarily improve the Faraday rotation signal, except if the background light from the medium (i.e. the continuum) is relatively bright. In this case, larger intensities reduced the impact of the background noise. In the proposed setup, the Faraday rotation derivative is proportional to the beam energy. So, the laser acts as an amplifier, allowing to increase the signal simply by increasing the beam energy. In real life situations, the spatial variation of the beam now adds to the Faraday rotation signal, as shown by Eq. \eqref{eq:general_polarization_derivative}. So it is imperative to smooth the beam to allow the signal. One possible method is simply to expand the beam, possibly beyond the optical system, to reduce unwanted gradients as much as possible near the optical axis. However, high frequency noise can be removed using spatial filtering, providing that the Faraday rotation signal varies much more slowly than the beam noise. 
\section*{Funding}
This research was supported by the NSF CAREER Award PHY-1943939 and by the Laboratory for Laser Energetics Horton Fellowships.
\section*{Disclosures}
The authors declare no conflicts of interest.
\section*{Data availability}
The data is available upon request to the first author.


\bibliography{biblio}

\end{document}